\documentclass[aps,twocolumn,preprintnumbers,floatfix]{revtex4-1}

\usepackage{graphicx, epsfig}
\usepackage{color}
\usepackage{mathrsfs}
\usepackage{amsmath}
\usepackage{hyperref}
\newcommand{\pprime}{{\prime\prime}}
\begin{document}
\title{An Exact Tunneling Solution in a Simple Realistic Landscape}
\author{Koushik Dutta}
\author{Pascal M.~Vaudrevange}
\author{Alexander Westphal}
\affiliation{Deutsches Elektronen-Synchrotron DESY, Theory Group, D-22603 Hamburg, Germany}
\preprint{DESY 11-028}
\begin{abstract}
  We present an analytical solution for the tunneling process in a
  piecewise linear and quadratic potential which does not make use of
  the thin-wall approximation. A quadratic potential allows for smooth
  attachment of various slopes exiting into the final minimum of a
  realistic potential. Our tunneling solution thus serves as a
  realistic approximation to situations such as populating a landscape
  of slow-roll inflationary regions by tunneling, and it is valid for
  all regimes of the barrier parameters. We shortly comment on the inclusion of gravity.
\end{abstract}
\date{February 23, 2011}
\maketitle

\begin{figure}
  \includegraphics[width=0.8\linewidth]{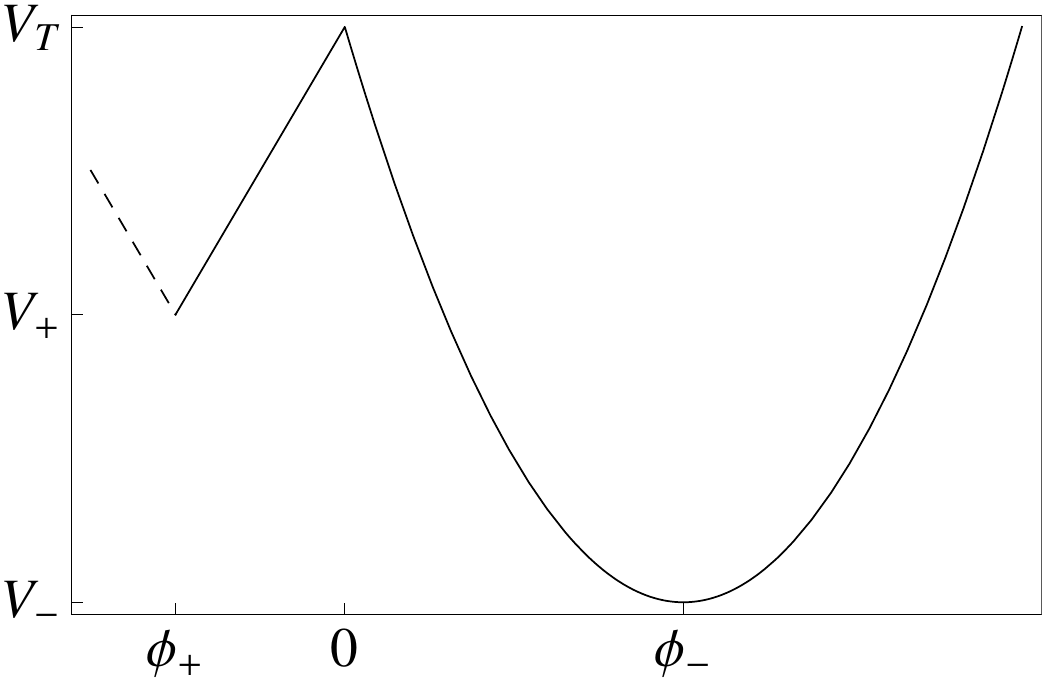}
  \caption{Schematic shape of the potential Eq.~(\ref{eq:potential})}
  \label{fig:potential_shape}
\end{figure}

Vacuum decay is one of the most drastic environmental changes in field theory, which often proceeds via nucleating a bubble of true vacuum inside a sea of false vacuum by a quantum mechanical tunneling event. Tunneling processes (in first order phase transitions) play a vital
role in many aspects of high-energy theory and cosmology. The groundwork for the computation of tunneling amplitudes was laid many years
ago by Coleman and de Luccia (CdL) \cite{Coleman:1977py, Coleman:1980aw}. Their method uses the Euclidean path integral to calculate what is now known as the CdL instanton, often appearing to be the stationary point of minimal action. The CdL analysis consist of a single real
scalar field in a potential with a false and a true vacuum located at the position of the corresponding minima of the scalar potential, 
$\phi_+$ and $\phi_-$ respectively, see
Figure~\ref{fig:potential_shape}. The tunneling probability per unit
volume $\Gamma/V = A e^{-B}\,,$ can be conveniently
expressed \footnote{$A$ is of order $m^4$ with $m$ being the characteristic scale of the problem~\cite{Callan:1977pt}, which we neglect here for simplicity.} in terms of the Euclidean action $S_E$ of the O(4)
symmetric so-called bounce solution $\phi_{{\text Bounce}}$ as $B =
S_E[\phi_{{\text Bounce}}] - S_E[\phi_+]\,$. In the interior of the nucleating bubble, the scalar field exits at some point on the slope towards the minimum, while outside of the bubble, beyond
the bubble wall, it still sits in the false minimum.

In general, the equation of motion for the bounce is difficult to
solve analytically. This lead to the development of the thin-wall
approximation \cite{Coleman:1977py}. In this limit of small energy
difference between the true and false vacuum, the bubble wall becomes
infinitely thin, and an approximate solution can be found.

To the best of our knowledge, an exact tunneling solution is only
known in the case of a piecewise linear potential
\cite{Duncan:1992ai}. In this article, we will derive the analytic tunneling
solution for the piecewise linear and quadratic potential
\begin{eqnarray}
  V(\phi)&=&\left\{\begin{array}{cl}
      V_L\equiv \lambda_+ \phi+V_{-}+\frac{1}{2}m^2\phi_-^2\,,&\phi<0\\
      V_R\equiv \frac{1}{2}m^2(\phi-\phi_-)^2+V_{-}\,,&\phi\ge0
    \end{array}\right. \label{eq:potential}\,,
\end{eqnarray}
(see Fig~\ref{fig:potential_shape}) which does not make use of the
thin-wall approximation, see also
recent work by \cite{Gen:1999gi}. However, one (non-differential) equation
remains to be solved either numerically or in the limits of a large
and small bubble radius. We show that our general analytical results
reduce to the thin-wall results and the results of
\cite{Duncan:1992ai} in the appropriate limits. For the most part of this work, we exclude
the effects of gravity and thus ignore the magnitude of the true
vacuum energy $V_{-}$. We will only offer some qualitative arguments
about the inclusion of gravity in the end, except for the thin-wall limit.

We expect this tunneling solution to be the simplest yet realistic approximation to an arbitrarily shaped barrier exiting smoothly into a true minimum. The reason is that a quadratic potential towards the true vacuum allows us to attach to its critical point smoothly a given shallowly sloped region of scalar potential containing a minimum.
This may be particularly relevant for studying the dynamics of populating a landscape of slow-roll inflationary slopes via tunneling from some false vacuum. Any discussion of the relative prevalence of different classes of inflationary models in a candidate fundamental theory such as string theory will necessarily have to include the discussion of the bias incurred by population via tunneling.

In the absence of gravity, the Euclidean action for a single scalar
field with potential energy $V(\phi)$ is given by
\begin{eqnarray}\label{eq:full:S_E}
  S_E&=&2\pi^2\int_{0}^{\infty}\!dr\,r^3\left(\frac{1}{2}\phi^{\prime^2}+V(\phi)\right)\,,
\end{eqnarray}
and the bounce solution $\phi(r)$ is determined by
solving the Euclidean $O(4)$ symmetric equations of motion
\begin{eqnarray} \label{eq:eom}
  \phi^\pprime(r)+\frac{3}{r}\phi^\prime(r)&=&\partial_\phi V\,,
\end{eqnarray}
where $\phi^\prime\equiv\partial_r\phi$ and the potential for our case is given by
Eq.~(\ref{eq:potential}).

\begin{figure}
  \includegraphics[width=0.8\linewidth]{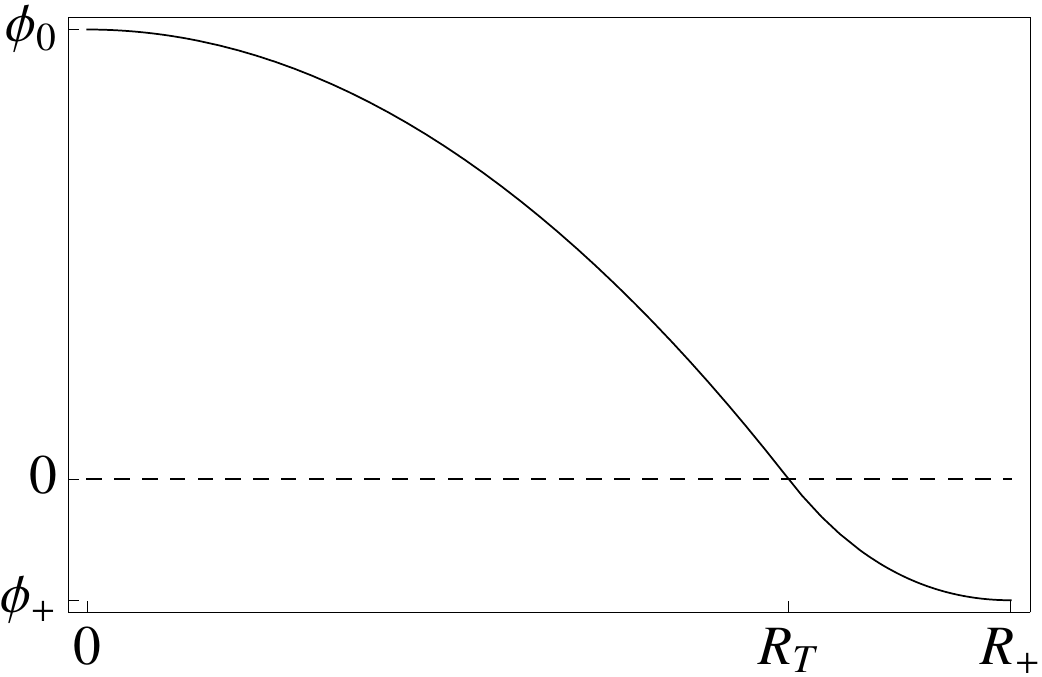}
  \caption{The bounce solution for $\alpha=10^{-5}, \Delta=10^{-5},
    m=10^{-6}, \phi_-=1$.The field sits at some value
    $\phi=\phi_0=0.0037$ inside the bubble at r = 0. At the bubble
    radius, the field crosses through the maximum locus of the
    potential at $\phi(R_T)=0$. Well outside the bubble, the field
    sits in the false vacuum $\phi(R_+)=\phi_+=-0.0001$. For the definitions of $\alpha$, $\Delta$ see the text below Eq.~\eqref{eq:full:B:exact}.}\vspace*{-3ex}
  \label{fig:BounceSolution}
\end{figure}

Initially, the field is sitting in the false vacuum at $\phi_+<0$,
kept in place by a linear potential piece (shown as a dashed line in Fig~\ref{fig:potential_shape})
attached to the left of the false minimum. Its slope is irrelevant in
the following computation as long as it classically stabilizes the
field in the false minimum. 
To start with, we first depict the numerically solved bounce solution 
schematically in Fig~\ref{fig:BounceSolution}. The
field sits at some value $\phi_0$ (not necessarily close to the true
vacuum $\phi_-=1$ in this case but to the right of the maximum of the
potential) inside the bubble at $r=0$. At the bubble radius $R_T$, the
field crosses through the maximum of the potential at $\phi(R_T)=0$.
Well outside the bubble, the field sits in the false vacuum
$\phi(R_+)=\phi_+=-0.0001$. 

The solutions $\phi_L$ and $\phi_R$ in the left  and right part of the
potential, respectively, have to fulfill the following boundary conditions: The
bubble nucleates at rest $\phi_R(0) = \phi_0>0, \phi_R^\prime(0)=0\,.$
Outside of the bubble wall at finite $r> R_{+}$, the field sits at
rest in the false vacuum $i.e$ $\phi_L(R_+) = \phi_+,
\phi_L^\prime(R_+)=0$. Solving Eq.~(\ref{eq:eom}) with these boundary conditions for both parts
of the potential, we find
\begin{eqnarray}\label{eq:full:phi_sols}
  \phi_L&=&\phi_++\frac{\lambda_+}{8r^2}\left(r^2-R_+^2\right)^2\,,\\
  \phi_R&=&\phi_-+2(\phi_0-\phi_-)\frac{I_1(mr)}{mr}\,,
\end{eqnarray}
where $I_1(z)$ is the modified Bessel function of the first kind (see
also equation (3.14) in \cite{Coleman:1977py}).

In order to determine the constants $\phi_0, R_+$, and $R_T$, we use
that the field configurations need to match smoothly at the bubble
radius $R_T$ with $\phi_L(R_T) = \phi_R(R_T) = 0$ and
$\phi_L^\prime(R_T) = \phi_R^\prime(R_T)$. Thus we need to solve
\begin{eqnarray}\label{eq:full:matching_conditions}
 (R_+^2-R_T^2)^2&=&-\frac{8R_T^2\phi_+}{\lambda_+}\;,\; I_1(mR_T)=\frac{\phi_-\,mR_T}{2(\phi_- - \phi_0)} \;,\nonumber\\
 R_+^4-R_T^4&=&\frac{8R_T^2}{\lambda_+}(\phi_--\phi_0)I_2(m R_T)\,.
\end{eqnarray}
Calculating the Euclidean action Eq.~(\ref{eq:full:S_E}) for the solutions
Eq.~(\ref{eq:full:phi_sols}), we can express both $\phi_0$ and $R_+$ in terms of
$R_T$ to obtain
\begin{eqnarray}\label{eq:full:B:exact}
  B&=&2\pi^2\phi_-^2R_T^2[\alpha^2 +\frac{1}{2}\left(\frac{4}{3}\alpha\sqrt{\Delta}+\frac{I_2(mR_T)}{I_1(mR_T)}\right)mR_T\nonumber\\
  &&-\frac{(1 - \Delta)}{8}m^2R_T^2 ],
\end{eqnarray}
where we have introduced $\alpha = -\phi_+/\phi_- > 0$, and $\Delta =
(-2\lambda_+\phi_+)/(m^2\phi_-^2)$ as a measure of the height of
the potential barrier with values $0<\Delta<1$.

In order to find $R_T$, we combine Eq.~(\ref{eq:full:matching_conditions})
to get
\begin{eqnarray}\label{eq:full:RtEq}
  2\alpha+\sqrt{\Delta}m R_T&=&m R_T \frac{I_2(m R_T)}{I_1(m R_T)}\,,
\end{eqnarray}
which can be solved numerically. However, it is much more instructive
to examine it in the limits of large and small $mR_T$. But first we
should briefly note that Eq.~(\ref{eq:full:RtEq}) can be used to
remove the Bessel functions from Eq.~(\ref{eq:full:B:exact}).
We emphasize that the above expression for $B$ now depends on the potential parameters and the only unknown quantity $R_T$, which is fully determined in terms of  $\alpha,\Delta$ by an algebraic Eq.~(\ref{eq:full:RtEq}). This is one of our central results.

Now we turn towards the two approximate solutions of our results. We start by taking the limit $m R_T \gg 1$, and in this limit, 
Eq.~(\ref{eq:full:RtEq}) 
can be solved as 
\begin{eqnarray}\label{eq:largemRt:RtFull}
 m R_T&=&\frac{3+4\alpha}{2}\frac{1}{1-\sqrt{\Delta}}\,.
\end{eqnarray}
We note that for all allowed values of $\alpha$ and $\Delta$, $m
R_T > \frac{3}{2}$. In particular, we find that $mR_T$ can be large either for the
thin-wall limit $\Delta\approx1$, or 
$\Delta<1$, but $\alpha>\Delta$ (see Eq.~(\ref{eq:separatrix})). For the latter case, the potential minima are separated by sizable distances in the field space, as well as the potential energy between the false and true vacuum. Therefore, the large $m R_T$ limit
encompasses more than the just thin-wall solution, see Table~\ref{Tab.1}. For small differences in the vacuum energy
$\epsilon 
=(1-\Delta)\frac{1}{2}m^2\phi_-^2$, in leading order of $\epsilon$,
\begin{eqnarray}\label{eq:largemRt:RtApprox}
 m R_T& \approx&\frac{(3+4\alpha)m^2\phi_-^2}{2\epsilon}\,,
\end{eqnarray}
which we will find to be identical to the result from using the
thin-wall formalism. This is in accord with expectations from the thin-wall formalism: The
radius of the nucleating bubble grows toward $\infty$ as the minima
become degenerate. Also, looking at the limit of the matching
condition $\phi_R=0$ in Eq.~(\ref{eq:full:matching_conditions})
\begin{eqnarray}
  \phi_0&=&\phi_-\left(1- \sqrt{\frac{\pi}{2}} (mR_T)^{3/2}e^{-mR_T}\right)\,,
\end{eqnarray}
it is clear that the true vacuum bubble nucleates close to the minimum
of the quadratic part of the potential, $\phi_0\sim\phi_{-}$,
for large values of $m R_T$.  Plugging in the expression for $R_T$ from
Eq.~(\ref{eq:largemRt:RtFull}) yields an unwieldy result for $B$.
Thus we only display the expression for $B$ in the
$\epsilon\rightarrow0$ limit
\begin{eqnarray}
  B&\approx&\frac{\pi^2}{96 \epsilon^3}m^4\phi_-^8\left(3+4\alpha\right)^4\,.\label{eq:largemRt:B}
\end{eqnarray}
Finally, we give an expression of $B$ in terms of $\phi_0$
\begin{eqnarray}
  B&\approx&\frac{\pi^2\phi_-^2}{12m^2} \,(4\alpha+3)\cdot\ln^3\hspace{-0.3ex}\Big(1-\frac{\phi_0}{\phi_-}\Big)\,.\label{eq:smallmRt2:B}
\end{eqnarray}
In summary, in this limit, when either the thin-wall approximations are valid or the minima are separated far away from each other, a large bubble nucleates close to the true vacuum.   

For the sake of comparison, we shall now give the essential results of the thin-wall calculation, where the nucleation rate is given by $B =
\frac{27\pi^2S_1^4}{2\epsilon^3}$ and for the potential
Eq.~(\ref{eq:potential}) we can compute $S_1$ directly, and in the small $\epsilon$ limit it is 
\begin{equation}
S_1 = -\frac{m}{2}\left(1 + \frac{4}{3} \alpha \right) \phi_{-}^2.
\end{equation}
Plugging this into the expression for $B$ 
we find the same expression as
Eq.~(\ref{eq:largemRt:B}). The same way, we get $R_T=3S_1/\epsilon$ which,
after plugging in $S_1$ and expanding in $\epsilon$, agrees with
\eqref{eq:largemRt:RtApprox}.

We will now proceed to the opposite case of taking the limit of small $m R_T \ll 1$. To second order in $mR_T$, Eq.~(\ref{eq:full:RtEq}) is solved by
\begin{eqnarray}\label{eq:smallmRt2:Rt}
  mR_T&=&2\left(\sqrt{\Delta}+\sqrt{2\alpha+\Delta}\right)\,,
\end{eqnarray}
where we discarded the negative solution. Now $mR_T<\mu_T$ can be 
equivalently casted as $ (0<\alpha < \mu_T^2/8) \wedge (0<\Delta < (\mu_T^2-8\alpha)^2/16\mu_T^2\,.$ Therefore small $m R_T$ limit corresponds to the small values of $\alpha$ and $\Delta$, where potential minima are closely spaced in the field space, but the potential difference between the false and the true minima is considerably large. In this limit, the exit point of the bubble can be conveniently expressed as 
\begin{equation}
\phi_{0} = \phi_{-}\left( 1 + \frac{8}{m R_T}\right)^{-1},
\end{equation}
and it shows that the bubble nucleates close to the tip of potential barrier which is intuitively understandable from the form the potential in this particular limit. 
We can express $B$ for small $mR_T$ in terms of $\alpha, \Delta$
as
\begin{eqnarray}\label{eq:smallmRt2:B:Rt}
  B&=&\frac{16\pi^2}{3}\frac{\phi_-^2}{m^2}\left(2\alpha+\Delta\right)\\
  &&\times\left[(\alpha+\Delta)^2+\Delta^2\left(1+\frac{2\alpha}{\Delta}\right)\left(1+\sqrt{1+\frac{2\alpha}{\Delta}}\right)\right].\nonumber
\end{eqnarray}
Again, we give an expression of $B$ in terms of $\phi_0$
\begin{eqnarray}
  B\approx \frac{16\pi^2\phi_-^2}{m^2}\left[\alpha^2\frac{\phi_0}{\phi_-}+\frac{4\alpha\sqrt{2\Delta}}{3}\left(\frac{\phi_0}{\phi_-}\right)^{\hspace*{-0.5ex}3/2}\hspace*{-2ex}+\Delta\left(\frac{\phi_0}{\phi_-}\right)^{2}\right]\label{eq:smallmRt2:B:phi0}
\end{eqnarray}

In the limit where the quadratic part of the potential can be regarded
as almost flat, the solution we found for small $mR_T$ should agree
with the results of a piecewise linear potential as studied by
\cite{Duncan:1992ai}. This is the case for the limit $\Delta\ll1$.
Taking the slope of the quadratic potential to be
$\lambda_-=m^2\phi_-$, the parameter $c$ defined by
\cite{Duncan:1992ai} becomes $c=2\alpha/\Delta$. The value of
the tunneling radius $m R_T$ found by \cite{Duncan:1992ai} is given by
\begin{eqnarray}
  m R_T=\frac{4\alpha}{\sqrt{2\alpha+\Delta}-\sqrt{\Delta}}
\end{eqnarray}
in agreement with the corresponding expansion of
 Eq.~(\ref{eq:smallmRt2:Rt}). Expanding the expression for $B$ in
Eq.~(\ref{eq:full:B:exact}), we find perfect agreement with $B$ from
\cite{Duncan:1992ai}.
\begin{figure*}[ht!]
  a)\includegraphics[width=0.31\linewidth]{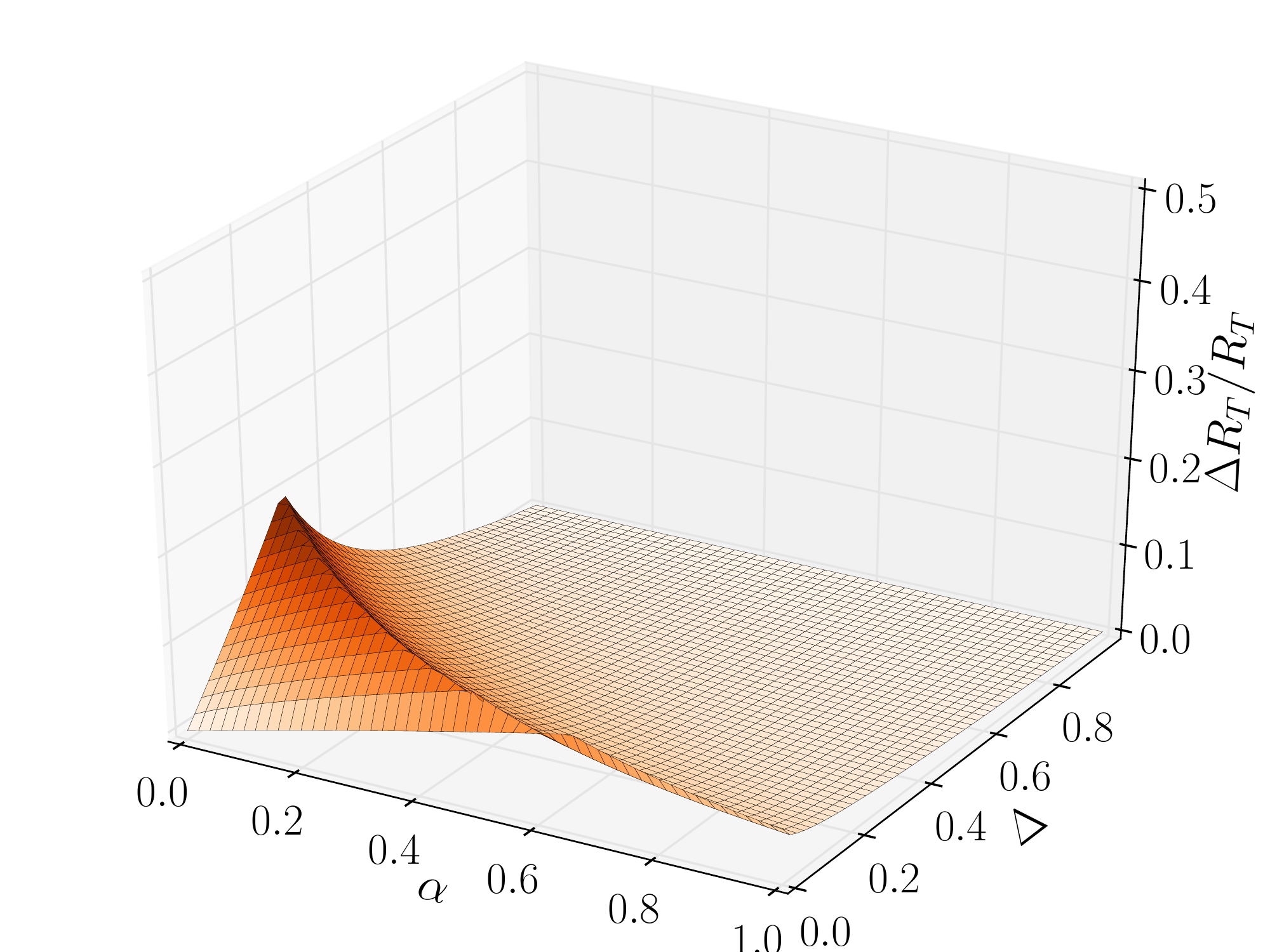}
  b)\includegraphics[width=0.31\linewidth]{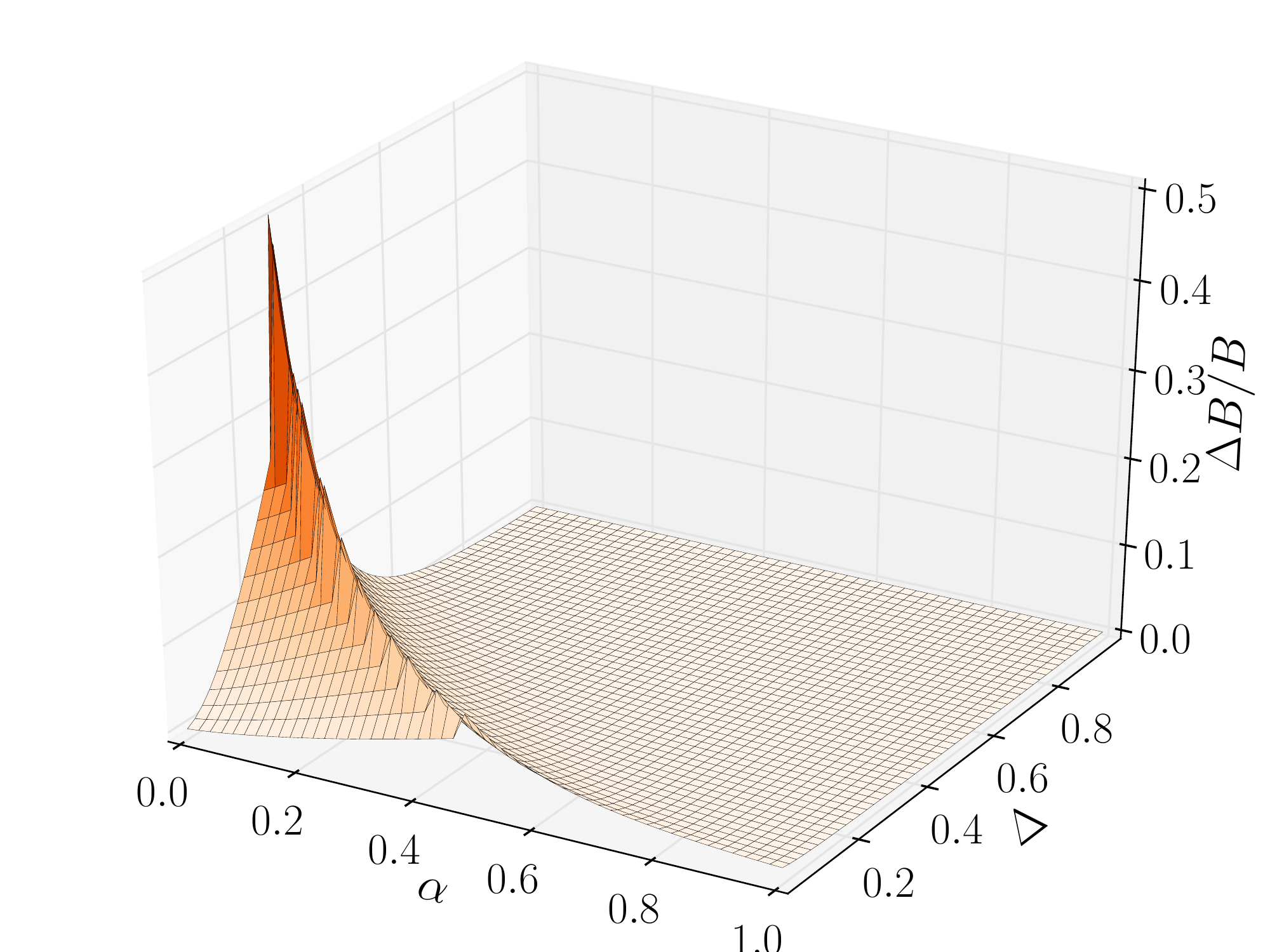}
  c)\includegraphics[width=0.32\linewidth]{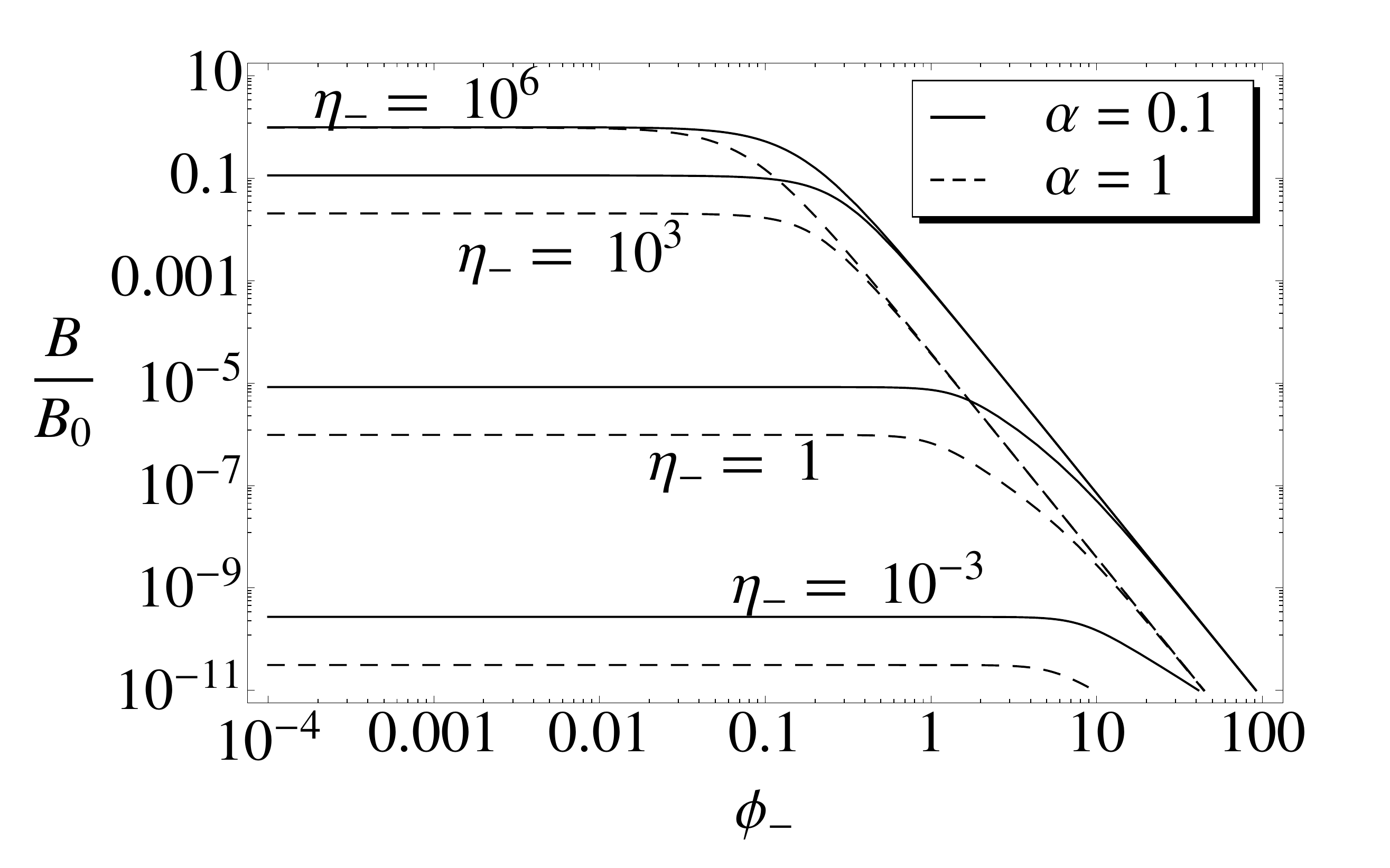}
  \vspace*{-3ex}\caption{The smaller of the relative error between the numerical
    value of $R_T$ (panel a) and $B$ (panel b) and its small or large
    $mR_T$ limit, Min$(|R_T^N-R_T^L|/R_T^N,
    |R_T^N-R_T^S|/R_T^N)$.  The relative error is always
    smaller than $10\%$ for $R_T$ and $50\%$ for $B$. c)
    $B/B_0$ of the tunneling rate for $\Delta = 0.95$ in the presence
    of gravity as a function of $\phi_{-}$, with $\eta_- \equiv
    m^2/V_{-}$. } \vspace*{-3ex}
  \label{fig:inclusion_of_gravity:BB0overRt}
  \label{fig:RelErrorB}
  \label{fig:RelErrormRt}
\end{figure*}

\begin{table}[t!]
  \begin{tabular}{lllll}
    & $mR_T$ & $B_{exact}$ & $B_{DJ}$ & $B_{thin-wall}$\\
    \hline
    $\alpha=0.01$ &  0.6 & 0.0023  & 0.0022  &  72.4  \\
    $\alpha=0.1$ & 1.2  & 0.3  & 0.04 &  113.3 \\
    $\alpha=0.5$ & 2.5  & 23.9  & 0.6  & 529.8   \\
  \end{tabular}
  \caption{The mismatch of action: the linearized result $B_{DJ}$
    of~\cite{Duncan:1992ai} compared to the thin-wall result $B_{thin-wall}$
    and the exact result found here. $B$ is given in units of 
    $\phi_-^2/m^2$ and all values are quoted for $\Delta=0.01$. 
    Clearly, there are regimes with $mR_T={\cal O}(1)$ where
    neither the linearized treatment of~\cite{Duncan:1992ai} 
    nor the thin-wall approximation are sufficient.}
  \label{Tab.1}
\end{table}

Let us recall what has been done so far. We reduced the computation of the
tunneling amplitude for tunneling from the linear to the quadratic
part of the potential Eq.~(\ref{eq:potential}) to solving an exact Eq.~(\ref{eq:full:RtEq}). We provided approximate solutions for
the limits of large and small tunnel radius $mR_T^{L,S}$.
Surprisingly, the smaller of the relative error between either the
numerical result $R_T^N$ and $R_T^L$ or $R_T^S$ is always smaller than
$10\%$, see Fig.~\ref{fig:RelErrormRt} (a). This then defines globally the approximate solution $mR_T$
\begin{eqnarray}
  m R_T=\left\{\begin{array}{cl}
      2\sqrt{\Delta}\left(1+\sqrt{1+\frac{2\alpha}{\Delta}}\right),& \Delta<(0.8\alpha-0.5)^2\\ & \\
      \frac{3+4\alpha}{2}\frac{1}{1-\sqrt{\Delta}}\;,&(0.8\alpha-0.5)^2<\Delta
      \end{array}
    \right.\label{eq:separatrix}
\end{eqnarray}
In other words, we succeeded
in computing the tunneling amplitude analytically in a non thin-wall solution with
error better than $50\%$, see Fig.~\ref{fig:RelErrormRt} (b).

Finally we shall comment on the effects of gravity on the tunneling
process. In the thin-wall limit \cite{Coleman:1980aw} studied the case where either true or
false vacuum are at zero energy, whereas \cite{Parke:1982pm} examined
this issue for arbitrary values of the false and true vacuum energy. Among other
things, their work suggests the following intuitive idea. When the
true vacuum is also in de-Sitter space, due to the ``pull'' on the
bubble both from the inside and the outside of the wall, the
nucleation rate should be enhanced compared to the flat space limit.
We can see this explicitly in the thin-wall approximation, {\it i.e.}
large $mR_T$ and $1-\Delta\ll 1$, where the effect of gravity is well
understood. Following \cite{Parke:1982pm}, we compute the effect of
gravity on the exponent $B$ of the tunnel rate, see
Fig.~\ref{fig:inclusion_of_gravity:BB0overRt} (c).

For sufficiently sub-Planckian values of $\phi_{-}$ we always have a
constant asymptotic value for $B/B_0$, where $B_{0}$ is without
gravity. In the case of small $\eta_- \equiv m^2/V_{-}$ ($i.e$ a very
flat quadratic part) the ratio $B/B_0$ is very small, giving a large
gravitational enhancement of tunneling. Contrary, for a steep
potential with large $\eta_-$ the tunneling probability is less
enhanced. Increasing $\alpha$ suppresses $B$ further and thus enhances
tunneling.  This fits with the notion, that the gravitational
correction is more important the thicker the barrier is in the
field space, since for a given $\phi_-$ the barrier thickness
increases with increasing $\alpha$. Note, that for $\alpha\gg 1$ and $\eta_-\ll1$ there is also Hawking-Moss tunneling possible~\cite{Hawking:1981fz,Starobinsky:1986fx,Linde:1991sk}.


In this work we have discussed quantum tunneling in field theory in a piecewise
linear and quadratic scalar potential with a a false vacuum and a true
vacuum. Such a potential is arguably the most simple yet realistic approximation
to an arbitrarily shaped barrier exiting smoothly into a true minimum. The reason is that a quadratic potential allows us to attach to its critical point smoothly a given shallowly
sloped region of scalar potential containing a minimum. Our result
gives the tunneling rate in this situation exactly. Further, in the
appropriate limits our result reduces to either the thin-wall result
or the known result for piecewise linear potentials
by~\cite{Duncan:1992ai}. However, there is a large region of barrier
shape parameter space where neither of them is a good approximation to
the full solution given here. The inclusion of further effects from approximating a given potential to higher than quadratic order, as well as a detailed incorporation of gravity in the non-thin-wall regime we leave as a topic for future work. Let us note in passing that in the context of meta-stable supersymmetry breaking vacua in gauge theories~\cite{Intriligator:2006dd}, choosing $N_f=3 N_c/2$ flavors can imply a barrier shape with $\alpha\ll\Delta\simeq 2/3$. For $\alpha=0.1$ this gives $mR_T\simeq 9$ which shows the linearized result of~\cite{Duncan:1992ai} with $B_{DJ}\simeq2.5\times 10^2$ overestimating the tunneling rate by an order of magnitude compared to our exact result $B\simeq 1.4\times 10^3$. Finally, we expect one relevant application of our results to be the
dynamics of populating a landscape of slow-roll inflationary
slopes via
tunneling from some false vacuum. For this purpose, we have given the
instanton action for the tunneling also as a function of $\phi_0$, as
the exit from the tunneling $\phi=\phi_0,\dot\phi_0=0$ forms
the initial conditions for  slow-roll inflation.

\noindent {\bf Note added:} While this paper was being finished, we became aware of~\cite{Pastras:2011zr}, which overlaps with our results.

\noindent {\bf Acknowledgements:} We are grateful to A.~Linde for enlightening comments. This work was supported by the Impuls und
Vernetzungsfond of the Helmholtz Association of German Research
Centres under grant HZ-NG-603, and German Science Foundation (DFG) within the 
Collaborative Research Center 676 ÔParticles, Strings and the Early UniverseÕ.

\bibliographystyle{apsrev4-1}
\bibliography{tunneling_linear_quadratic_v1}
\end{document}